\documentstyle[12pt]{article}
\def\MSbar{\relax\ifmmode\overline{\rm MS}\else{$\overline
{\rm MS}${ }}\fi}
\def \as{\relax\ifmmode\alpha_s\else{$\alpha_s${ }}\fi}
\def\abar{\relax\ifmmode{\bar{a}}\else{$\bar{a}${ }}\fi}
\def\albar{\relax\ifmmode{\bar{\alpha}}\else{$\bar{\alpha}${ }}\fi}
\def\albars{\relax\ifmmode{\bar{\alpha}_s}\else{$\bar{\alpha}_s${ }}\fi}
\def \asQ{\relax\ifmmode\bar  \alpha_s(Q)\else{$\bar \alpha_s(Q)${ }}\fi}
\def \asZ{\relax\ifmmode\bar  \alpha_s(M_Z)\else{$\bar \alpha_s(M_Z)${
}}\fi}
\def \asQm{\relax\ifmmode\bar \alpha_s(Q,m)\else{$\bar \alpha_s(Q,m)${ }}
\fi}
\def \asQM{\relax\ifmmode\bar \alpha_s(Q,M)\else{$\bar \alpha_s(Q,M)${ }}
\fi}
\def\cM{{\cal{M}}}

\def\gl{\tilde{g}}

\newcommand{\la}{\label}     
\def\ie{\hbox{\it i.e.}{}}  
\def\eg{\hbox{\it e.g.}{}}   
\def\beqlab#1{\begin{equation}\label{#1}}
\newcommand{\Ds}{\displaystyle}  
\def\pr#1#2#3{{\em Phys. Rev.} {\bf D#1} (19#3) #2}
\def\np#1#2#3{{\em Nucl. Phys.} {\bf B#1} (19#3) #2}

\def\pl#1#2#3{{\em Phys. Lett.} {\bf B#1} (19#3) #2}

\newcommand{\bd}{\begin{displaymath}} \newcommand{\ed}{\end{displaymath}}
\newcommand{\ba}{\begin{eqnarray}}    \newcommand{\ea}{\end{eqnarray}}
\newcommand{\beq}{\begin{equation}} \newcommand{\eeq}{\end{equation}}
\newcommand{\baa}{\begin{array}{lll}} \newcommand{\eaa}{\end{array}}

\begin{document} 
\begin{center}
{\large\bf  Continuous mass-dependent analysis \\
 of the non--singlet $xF_3$  CCFR  data} \\
\vspace{10mm}
D.V. Shirkov,
\footnote{E-mail: shirkov@th-head.jinr.dubna.su}
A.V. Sidorov
\footnote{E-mail: sidorov@thsun1.jinr.dubna.su}
 and S.V. Mikhailov
\footnote{E-mail: mikhs@thsun1.jinr.dubna.su}\\

{\it Bogoliubov Laboratory, JINR, Dubna, Russia} \\
\end{center}
\abstract{We consider the issue of an accurate description of the
evolution
of the non-singlet structure function moments $M_n(Q)$ near heavy quark
threshold. To this aim we propose a simple modification of the standard
massless \MSbar scheme approach to the next-to-leading QCD analysis of
DIS data. We apply it to the processing of the modern CCFR'97 data
for $xF_3$ structure function and extract
the value of
 $$\alpha_s(M_z) \approx 0.122 \pm 0.004$$
We check also the consistency of light gluino hypothesis with
CCFR'97 data.}


\section{Introduction}
An important means of verification of the validity of pQCD
(that is, of perturbative QCD "improved" by the RG summation) is an
analysis of deep inelastic scattering (DIS) data. To interpret these
data within pQCD, one should pay credit to a number of subtle physical
effects: contributions of high twists, nuclear effects, high--order
(three-loop) corrections and the influence of thresholds of heavy
particles. All the introduced corrections are roughly
of the same order of magnitude.

This paper is devoted to the problem of influence of the thresholds
of heavy quark (HQ) on the pQCD analysis of DIS data that includes,
in particular, the evolution of the strong coupling constant
$\albars(Q)$. Modern estimates performed in ~\cite{yuri,mikh,chyla}
have revealed a significant role of threshold effects in the
 $\albars(Q)$ evolution when the DIS data lie close to
 the position of "Euclidean--reflected" threshold of heavy particles.
The corresponding corrections to $\as(M_Z)$ can reach several per
cent, \ie , they are of the order of the three--loop ~\cite{chyla},
for the recent results see ~\cite{kataev},
and nuclear effects ~\cite{ST95} on $\as(M_Z)$ .

A common algorithm for the renormalization--group (RG) resummation
is based upon beta--function $\beta(\as)$ and anomalous dimensions
$\gamma(\as)$ calculation and the RG differential equations integration
performed within the \MSbar renormalization scheme. However, the
widespread
massless \MSbar scheme fails to describe the data near thresholds
of heavy particles -- $b, c$  quarks and, maybe, light superpartners
~\cite{mikh}.

An appropriate procedure for the inclusion of threshold effects into the
$Q^2$--dependence of $\albars(Q)$ in the framework of the massless
\MSbar scheme was proposed more than 10 years ago ~\cite{bern,marc} :
transition from the region with a given number of flavors $f$
described by massless $\albars(Q;f)$ to the
next one with $f + 1$ (``transition across the $M_{f+1}$ threshold")
is realized here with the use of the so--called ``matching relation"
for $\asQ$ ~\cite{marc}.  The latter may be considered as the
continuity condition for $\albars(Q)$ on (every) HQ mass
\beq \la{marc}
 \albars(Q=M_{f+1}; f)=\albars(Q=M_{f+1}; f+1)~ \eeq
that provides an accurate $\albars(Q)$--evolution description for $Q$
values not close to the threshold region. The condition (\ref{marc})
is used up to the three--loop level; the other version of the matching
can be found in ~\cite{bern}.

     One needs also one more element, the matching procedure for the
evolution of the structure function moment $M_n(Q,m)$. The corresponding
expressions for anomalous dimensions $\gamma_{(i)}(n;f)$ are well
known in the \MSbar scheme for a fixed $f$ value (see, e.g.,~\cite{GA})
but until now there is no recipe for obtaining a continuous
interpolation across the HQ threshold for the moment evolution.

In this paper, we are going to focus just on this aspect of the
problem: how does the HQ threshold influence the evolution of the DIS
structure function? We will examine only non-singlet processes of DIS
so as to pass over the delicate problem of modification of the operator
product expansion in DIS through introducing a new scale, the mass of a
heavy parton (for discussion, see ~\cite{gotts}).

To solve the problem we propose a rather simple modification of the
massless
\MSbar scheme to take into account thresholds in analyzing the moments
of the DIS non-singlet structure function at the two-loop level.

   To simplify the exposition, we shall take advantage of the explicit
analytic mass-dependent RG--solution derived in ~\cite{tmf81} and
~\cite{np92}
that is expressed directly in terms of \asQm and $M_n(Q,m)$ perturbation
expansion coefficients. This allows us to avoid the use of
RG--generators, that are $\beta$ and $\gamma$--functions. In the next
section, we present smooth analytic expressions for the evolution
of $\asQm$ and $M_n(Q,m)$ at the 2--loop level based on this
mass--dependent RG formalism. We shall omit all theoretical and
technical details (they can be found in refs.~\cite{tmf81,np92,Brussel95})
 and write only final results.
In Sect.~3, we introduce the``spline-approximation" to describe the
two-loop level continuous moment evolution, and present there
another proof of the matching condition (\ref{marc}).
In Sect.~4, we describe briefly a method of analysis of the DIS data.
On its base  we carry out the fit of CCFR'97 Collab. data,
extract the parameter $\as$ and estimate the contribution of
threshold effects.
We discuss the consistency of MSSM light gluino existence with
CCFR'97 data
by using the spline--type evolution of $M_n(Q,m)$ in Sect.~5.

Throughout the paper we use the notation: $a = \as/4\pi,
~(\abar  = \albars/4\pi)$;
indices in brackets stand for the loop number, \eg , $\beta_{(\ell)}=
\beta_{\ell -1}$; instead of the structure function moments
$M_n(Q,m)$, we consider only its ``evolution part" $\cM_n(Q,m)$
(i.e., moments of the distribution function)
\beqlab{moment}
        M_n(Q,m)=C_n(a,Q,m) \cdot \cM_n(Q,m),
\eeq
where $C_n$ are moments of the coefficient function of a certain DIS
process (see, \eg, Ref.~\cite{tbook}).

\section{Mass-dependent RG solutions}
    In the massless case, the moment two-loop evolution is
 described by the expression
\beqlab{m2} \Ds
 \cM_n^{(2)}(Q)=\cM_n(\mu) \left(\frac{a}{\abar^{(2)}(Q)}\right)^
{d_n} \exp \left\{ \left[ a -\abar^{(2)}(Q)\right]
f_n\right\}~\eeq  
with the ~~  n u m e r i c a l ~~coefficients
\beq
\la{nMS}
  d_n=\frac{\gamma_0(n)}{\beta_0}~;~~ f_n =\frac{\beta_0\gamma_1(n) -
\beta_1\gamma_0(n)}{\beta_0^2}~; \eeq
 $$~\beta_0= \beta_{(1)}(f) = 11-\Delta\beta_{(1)}f~, \
\Delta\beta_{(1)} =2/3~;~~\beta_1 =\beta_{(2)}(f)=102-\Delta\beta_{(2)} f,
 \ \Delta\beta_{(2)}=38/3~. $$

  In the mass-dependent case, one should use instead of Eq.(\ref{m2}),
a bit more complicated expression \cite{tmf81} of the same structure
\beqlab{m-rg2}
\cM_n^{(2)}(Q)=\cM_n(\mu) \left( \frac{a}{\abar^{(2)}(Q,m)}\right)^
{D_n(Q,...)}\exp \left\{ \left[ a -\abar^{(2)}(Q,m)\right]
F_n(Q,...)\right\} \eeq
with the ~~ f u n c t i o n a l ~~coefficients $D_n, F_n$
\beqlab{d-n}
D_n(Q,m,\mu)=\frac{\Gamma_{(1)}(n,Q)}{A_{(1)}(Q, m, \mu)}~;
\eeq
\beqlab{fn-m} F_n(Q, m, \mu) = \frac{A_{(1)}(Q, m, \mu)\Gamma_{(2)}(n,Q)
-A_{(2)}(Q, m, \mu)\Gamma_{(1)}(n,Q)}{[A_{(1)}(Q, m, \mu)]^2}
 \eeq
and the two-loop running coupling \abar taken in the form
\beqlab{a-rg2}
\frac{1}{\abar^{(2)} (Q,m;a)}= \frac{1}{a}+A_{(1)}(Q, m, \mu)+
\frac{A_{(2)}(Q, m, \mu)}{A_{(1)}(Q, m, \mu)}\ln[1+
a A_{(1)}(Q, m, \mu)]~. \eeq

In the non-singlet case of DIS, \
$$\Gamma_{(1)}(n, m, Q)=\Gamma_{(1)}(n,l)=\gamma_{(1)}(n)l,
~l=\ln\left(Q^2/\mu^2\right)~,$$~
and the HQ-mass-dependent
$A_{(\ell)}$,~$\Gamma_{(\ell=2)}(n,Q)$ appearing  in Exp.
(\ref{d-n}-\ref{a-rg2}) are just perturbation expansion coefficients:
$$
\abar(Q,m,\mu;a)_{pert} = a-a^2A_{(1)}(Q, m, \mu)+a^3\left\{\left[
A_{(1)}(Q, ... )\right]^2- A_{(2)}(Q, m, \mu) \right\}~+ \ldots;
$$ 
\ba
\la{m-pt}
\left. \frac{\cM_n(Q_)}{\cM_n(Q_0)}\right|_{pert} = 1 + a
\Gamma_{(1)}(n,l) + \hspace{90mm}  \nonumber \\
\phantom{\frac{\cM_n(Q_)}{\cM_n(Q_0)}} + a^2\left\{\frac{\Gamma_{(1)}(n,l)
\left(\Gamma_{(1)}(n,l)-A_{(1)}(Q, m, \mu)\right)}{2}+
\Gamma_{(2)}(n, Q, m, \mu) \right\}~ +\ldots \quad
\ea
satisfying the normalization condition --
$A_{(\ell)}(Q=\mu) = \Gamma_{(\ell)}(n,Q=\mu) =0 $. \\
These coefficients consist of the usual massless part (for $f=3$) and
HQ-mass dependent contributions, \eg ,
\beqlab{a-1}
~A_{(1)}(Q, m, \mu) = \left( \frac{1}{\abar^{(1)}(Q,m,\mu;a)} -
\frac{1}{a}\right) =\beta_{(1)}(3)\/l-\Delta\beta_{(1)}
\sum\limits_{h}^{ }\left[I_1\left(\frac{Q^2}{m^2_h}\right)- C\right]~
\eeq
with summation over HQ's: $h\geq 4$.
Here $I_1$ is the one-loop fermion mass-dependent contribution, like
the polarization operator ~\cite{GePo} or the three--gluon vertex loop
~\cite{NaWe} subtracted at $Q^2=0$ and $C$ being some subtraction
scheme-dependent constant.

``Massive" RG solutions (\ref{m-rg2})
and (\ref{a-rg2}) possess several remarkable properties:  \par
\begin{itemize}
\item they are built up only of ``perturbative bricks", \ie,
loop-expansion coefficients $A_{(\ell)}$, $~\Gamma_{(\ell)}(n)$\/
(taken just in the form they appear in the perturbative
input) and ``contain no memory" about the
intermediate RG entities such as $\beta$ and $\gamma$ functions;
\item  in the massless case with pure logarithmic
 coefficients,  $A_{(k)}=\beta_{(k)}l, \
\Gamma_{(k)}(n)=\gamma_{(k)}(n)l$,  they precisely correspond to the
usual massless expressions, like Eq.(\ref{m2});
\item  being used in QCD, they smoothly interpolate across
heavy quark threshold between massless solutions with different
flavors numbers.
\end{itemize}
\section{Smooth schemes and MS massless schemes}

\subsection{Smooth mass-dependent scheme}
We have above considered general formulae to describe the
$\cM_n(Q)$-evolution including the threshold effects.
It is clear that the mass--dependent MOM schemes automatically provide
the most natural smooth description of thresholds. To use them
in the framework of leading order, one needs a mass-dependent expression
for
$I_1$ presented of Appendix A (for two different schemes). So, to perform
the one-loop evolution analysis of moments, one should substitute
Eq.~(\ref{exactH1}) or (\ref{a5}) in Appendix A into Eq.~(\ref{a-1}) and
then into Eq.~(\ref{a-rg2}) and Eq.~(\ref{d-n}), and use the approximation

\beqlab{m-rg1}
\cM_n^{(1)}(Q)=\cM_n(\mu) \left( \frac{a}{\abar^{(1)}(Q,...)}\right)^
{D_n(Q,...)}~. \eeq
We have performed the results of the fit of the ``old" CCFR data
~\cite{CCFR93} following this formula in ~\cite{SSM96}.

However, the MOM scheme meets tremendous calculational difficulties in
the next-to-leading order of pQCD.
Moreover, each of the expansion coefficients
$A_{(i)}$, $\Gamma_{(i)}$ (see Sec. 2) becomes gauge--dependent at the
two-loop level, which is not convenient.
These difficulties are absent in the widespread \MSbar scheme.
One can go far in loop calculations here (see ~\cite{LRV}), but the scheme
is not sensitive to the thresholds at all.
Below we suggest a practical compromise between
these different possibilities -- the ``spline" scheme. This scheme
possesses
both the sensitivity to thresholds and simplicity of the \MSbar procedure.
Nevertheless, the \MSbar scheme looks like a conventional standard for all
DIS calculations now. Therefore, one should recalculate the results
obtained
in other schemes to the \MSbar scheme at an appropriate number $f$.
We do not need recalculation for $\as^{\MSbar spl}(M_Z)$ because the
spline
and \MSbar (at $f=5$) schemes evidently coincide at $\mu=M_Z$ by
construction,
\ie, ~~$\as^{\MSbar spl}(M_Z)=\as^{\MSbar}(M_Z;f=5)$.

\subsection{\MSbar -- vulgate scheme }
Usually, to obtain the evolution law, one calculates numerical
expansion coefficients of generators $\beta(\as)$, $\gamma_n(\as)...$
 in \MSbar scheme and solves the massless RG equations. In the
solution, with all integration constants being omitted, one arrives
at the final procedure which we shall name the \MSbar --vulgate scheme.

The first recipe to include the threshold mass $M_f$ into the framework of
the \MSbar evolution  was formulated in Ref.~\cite{marc} as the ``matching
condition", Eq.~(\ref{marc}), for the coupling constant.
Now all measurements on a low scale $Q$ are usually interpreted
in terms of the $\alpha_s(M_Z)$ --  RG solution in a certain scheme, with
an appropriate matching of different numbers of active flavors
which evolve from the scale $Q$ to $M_Z$.
The matching condition (\ref{marc}) leads to a simple rule of including
next ``active flavors" $h$ into the evolution law 
\ba
\la{a-i}
A^{\MSbar}_{(i)}(l) = \beta_{(i)}(3)\/l- \Delta\beta_{(i)} \cdot h l \to
~A^{\MSbar spl}_{(i)}(Q, ...) =
\beta_{(i)}(3)\/l- \Delta\beta_{(i)} l^* ;\\
l^* = \sum\limits_{h}^{ }
\left[ \theta(Q^2-(M_{h(i)})^2)\ln(Q/M_{h(i)})^2 - ( Q \to \mu) \right]~
\nonumber
\ea
Nevertheless, the threshold value $M_{h(i)}$ does not follow from this
procedure and is left uncertain.

Note, the {\it spline-type} (in terms of the $\/l$-variable) expression
(\ref{a-i}) has an evident analogy with the approximation for the
mass-dependent
MOM scheme formulae for $A_{(i)}(Q,\mu)$ ($~\Gamma_{(i)}(n,Q,\mu)$)
with the structure
$A_{(i)}(Q, \mu) \sim \left(I_{(i)}(Q, m) - I_{(i)}(\mu, m)\right)$,
see, \eg, Eq.~(\ref{a-1}). The approximation being discussed needs an
asymptotic form of the mass-dependent calculation for the elements
$I_{(i)}(z = Q^2/m^2)$, \ie, only logarithmic and constant terms
$I_{(i)}(z) \to \ln(z) - c_i$ (see, \eg, Exp. (\ref{asym1})
and (\ref{a6}) in Appendix A). Based on this form one can construct a
simple
 ``pure log" ansatz for $I_{(i)}(Q,m)$:
$$ I_{(i)}(Q,m) \to I_{(i)}^{MOM spl}(Q,\tilde{M}_{h(i)})=
\theta(Q^2-\tilde{M}_{h(i)}^2)\ln\left(Q^2/\tilde{M}_{h(i)}^2\right); \ \
\tilde{M}_{h(i)} = m_h \exp(c_i/2). $$
This ansatz roughly imitates the ``decoupling" property of $I_{(i)}(Q,m)$
 at $Q < \tilde{M}_{h(i)}$ and provides its asymptotic form at
$Q > \tilde{M}_{h(i)}$. It leads to the approximation for $A_{(i)}(Q,
\mu)$:
\ba
\la{a1-spl}
A_{(i)}(Q, \mu) \to &A_{(i)}^{MOM spl}(l)& = \nonumber \\
&\beta_{(i)}(3)\/l& -\Delta\beta_{(i)}\sum\limits_{h}^{}
\left[\theta(Q^2-
\tilde{M}_{h(i)}^2)\ln\left(Q^2/\tilde{M}_{h(i)}^2\right) -
( Q \to \mu)\right],
\ea
where the threshold position $\tilde{M}_{h(i)}$ is determined by the
scheme
dependent constant $c_{(i)}$.

A certain  value of the threshold $M_{h(i)}$ in Exp.(\ref{a-i}) and
another
proof of the matching condition (\ref{marc}) for the \MSbar scheme can be
obtained by using, \eg, the ``three--step procedure" introduced in
\cite{mikh}. Let us review it briefly.

Well below the threshold, for $\mu \ll M$, one usually uses some effective
\MSbar scheme, say $\MSbar_1$, that does not take mass of a particle into
account.
Above the threshold, a new particle cannot be ignored, but when
$Q \gg M$, it can approximately be treated as massless within some
other $\MSbar_2$ scheme. How should the couplings $a_1(\mu)$ and
$a_2(Q)$ in these two MS schemes be related? The answer can be obtained
by the three step algorithm:

  (i) recalculating from the $\MSbar_1$ to MOM scheme at $q=\mu \ll M$
      to get $a^{MOM}(\mu)$;

  (ii) performing the RG evolution of $a^{MOM}$ up to $q=Q \gg M$ in the
        MOM scheme;

  (iii) recalculating  to the $\MSbar_2$ scheme, including the mass
         contribution, at $q=Q$.\\
The final result of these successive steps leads to the approximate
(due to power corrections) equality $M_h \approx m_h$ \cite{mikh} for
the threshold at the two--loop level.
We obtain just the same result as usually used for the matching condition
mentioned above with $M_h=m_h$.

Consequently, proceeding in this way, one {\bf must modify} the
perturbative expansion coefficients for $\cM_n$ ,\ie ,
$\Gamma_{(i)}(n,Q)$ in (\ref{m-pt}), in the same manner as the
expansion coefficients of the coupling constant
 $A_{(i)}=\beta_{i}\/l \to A^{\MSbar spl}_{(i)}$. For this aim we recall
the structure for $\gamma_{(2)}(n;f)$ in the framework of the \MSbar
scheme (for details see, \eg,~\cite{GA},~\cite{tbook})
\ba \la{g2}
~\gamma_{(2)}(n;h+3) = \gamma_{(2)}(n; 3)-
\Delta\beta_1 \cdot h\cdot \Delta\gamma_{(2)}(n); \\
~\Delta\gamma_{(2)}(n)= \frac{16}{9} \left\{10 S_1(n) -6 S_2(n) -
\frac{3}{4} - {11n^2+5n-3 \over n^2 (n+1)^2}\right\}; \ \ S_k(n)=
\sum_{j=1}^n \frac1{j^k}, \nonumber
\ea
where the first term $\gamma_{(2)}(n; 3)$ in the r.h.s. of Eq. (\ref{g2})
consists of the usual massless part and the parameter $h$ numbers here
heavy
flavors. It is known (see \eg ~\cite{mr85}) that $\gamma_{(2)}(n; f)$
contains the terms generated by the evolution of the coupling constant.
These terms naturally
appear in the calculation of two-loop diagrams for $\gamma_{(2)}(n; f)$,
they are proportional to the coefficient $\beta_{(1)}$ (see the second
term
in (\ref{g2})). Therefore in the \MSbar - expression for ~$\Gamma_{(2)}$
there appears a term proportional to the one-loop coefficient $A_{(1)}$:
\beq \la{G}
~\Gamma_{(2)}(n,Q;h+3) = \Gamma_{(2)}(n,Q; 3)-
\Delta\beta_1 (h \cdot \/l )\cdot \Delta\gamma_{(2)}(n); \eeq
$$~\Gamma_{(2)}(n,Q;3)=\gamma_{(2)}(n, 3)\/l. $$
The $h$-dependent part of the $A_{(1)}$--term is singled out
of Exp.(\ref{G}) in the form of
$\Delta\beta_1 (h \cdot \/l)$.    
To obtain the continuous coefficient $\Gamma_{(2)}^{\MSbar spl}(n, Q,
h+3)$,
one should substitute $A_{(1)} \to A^{\MSbar spl}_{(1)}$, \ie, \
~$(h \cdot\/l) \to \/l^*$ into the second term of the r.h.s. of Eq.
(\ref{G}),
according to the recipe (\ref{a-i}):
\ba
\la{Gspl}
~\Gamma_{(2)}(n, Q; h+3) \to
 ~\Gamma_{(2)}^{\MSbar spl}(n,Q;h+3) = \Gamma_{(2)}(n,Q;3)-
\Delta\beta_1 l^* \cdot \Delta\gamma_{(2)}(n).
\ea
Now we can get the complete evolution law by substituting (\ref{a-i}) and
(\ref{Gspl}) into formulae (\ref{a-rg2}) and (\ref{d-n}), (\ref{fn-m})
and then into the general formula (\ref{m-rg2}):
\beq
\la{last}
\frac{\cM_n^{(2)}(Q)}{\cM_n(\mu)}=
\left( \frac{a}{\abar^{(2)}_{\MSbar spl}(Q,M)}\right)^{D_n(Q,M)}
\exp \left\{ \left[ a -\abar^{(1)}_{\MSbar spl}(l)\right]
F_n^{\MSbar spl}(Q,M)\right\} .
\eeq
 Recent CCFR'97 Collab. experimental data on DIS ~\cite{prep1}
 are processed by this method
(taking also account of the one--loop coefficient function) in
the next section.

\section{The QCD fit of the $~xF_3~$ CCFR data}
\subsection{ Method of QCD Analysis }

In this section, we present the QCD analysis of the
CCFR'97 data \cite{prep1}. They are the most precise data on
the structure function
$~xF_3(x,Q^2)~$. This structure function is pure non-singlet and the
results
of analysis are independent of the assumption on the shape of gluons. To
analyze the data, the method of reconstruction of the structure functions
from their Mellin moments is used \cite{Kriv} . This method is based on
the
Jacobi - polynomial expansion of the structure functions.

Following the method
\cite{Kriv,Jacobi}, we can write
 the structure
function $~xF_3~$ in the form:
\begin{equation}
xF_{3}^{N_{max}}(x,Q^2)= x^{\alpha }(1-x)^{\beta}\sum_{n=
0}^{N_{max}}\Theta_n
 ^{\alpha , \beta}
(x)\sum_{j= 0}^{n}c_{j}^{(n)}{(\alpha ,\beta )}
M_{j+2}^{NS} \left ( Q^{2}\right ),   \\
\label{e7}
\end{equation}
where $~\Theta^{\alpha \beta}_{n}(x)~$ is a set of Jacobi polynomials and
$~c^{n}_{j}(\alpha,\beta)~$ are coefficients of their
 power expantions:
\begin{equation}
\Theta_{n} ^{\alpha , \beta}(x)=
\sum_{j= 0}^{n}c_{j}^{(n)}{(\alpha ,\beta )}x^j .
\label{e9}
\end{equation}

The quantities $N_{max},~ \alpha~$ and $~\beta~$ have to be chosen so
as to
achieve the fastest convergence of the series in the r.h.s. of
Eq.(\ref{e7})
and to reconstruct $~xF_3(x, Q^2)~$ with the accuracy required. Following
the results of \cite{Kriv} we have fixed
the parameters -- $~\alpha =  0.12~, ~\beta =  2.0~$ and $~N_{max} =
12~$.
These numbers guarantee an accuracy better than $~10^{-3}~$.

Finally, we have to parameterize the structure function $~xF_3(x,Q^2)~$
at some fixed value of $~Q^2 =  Q^2_{0}~$. We choose  $~xF_3(x,Q^2)~$ in a
little bit more general form as compared to \cite{KaSi}, where the same
data have been analyzed within QCD in terms of $\Lambda^{\MSbar}_{(4)}$
without thresholds effects:
\begin{equation}
  xF_{3}(x,Q_0^2)= Ax^{B}(1-x)^{C}~(1+\gamma~x).
\label{e10}
\end{equation}
Here A, B, C, $\gamma$ and  $~\alpha_{0}=\albars(Q_0)~$
are free parameters to be determined by the fit.

To avoid the influence of higher--twist effects,
 we have used only the experimental points in the plane $~(x,Q^2)~$
with $~3 < Q^2\leq 200~(GeV/c)^2~$  and $~0.01\leq x \leq 0.75~$.
The effect of target--mass corrections in $xF_3$ is taken
into account to order $M^2/Q^2$ \cite{GEORGI}.\\

\subsection{ Results of Fit and Discussion }
Here we present the results of processing the CCFR'97 data obtained
in the framework of two different approaches:

First, we have used the massless $f$-fixed $\MSbar$ - scheme approach
based on the two-loop evolution formula (\ref{m2}). The corresponding
results for $\alpha_0(f)$ are collected in the up-parts of the
 of Table 1 for every $Q_0^2$.

The second approach is based on mass-dependent evolution Eq.(\ref{m-rg2}),
for this formula we adapt the ``spline" approximation (\ref{last}). These
results for $\as^{spl}$ are presented in the down-parts of Table 1 for
every
$Q_0^2$.

The results in Table 1
are completed by the results of evolution of $\alpha_0(f)$ and
$\as^{spl}$
to the point $Q=M_Z$ with appropriate matching (\ref{marc}) of different
numbers of active flavors (the last column -- $\as(M_Z)$; see, for
comparison with other estimations, review ~\cite{Beth95}). Repeating
the fit procedure for different values of $Q_0^2=3,\ 30,\ 200 \
\mbox{GeV}^2$ we
have obtained the experimental dependence of $\alpha_0(f)$ on the
momentum transfer. In all fits, only statistical errors are taken into
account. Here we make some comments on the fit results.
\begin{itemize}
\item To demonstrate the sensitivity of $\alpha_0(f)$ on the $f$,
the results of the fit are shown for different $f$ for each $Q_0^2$.
The largest difference between the values of $\alpha_0(f)$ for different
$f$ is about 8\% in the case of $Q^2_0$ close to kinematical boundaries
 $Q^2_0=3 \mbox{ and~} 200~GeV^2$.
This difference reduces to $2.5\% - 4\%$ variations for $\as(M_Z;f)$.
There are opposite relations for $\alpha_0(f)$  for these two points:
$\alpha_0(3)>\alpha_0(4)>\alpha_0(5)$ for $Q^2_0=3~GeV^2$ and
$\alpha_0(3)<\alpha_0(4)<\alpha_0(5)$ for $Q^2_0=30,~200~GeV^2$.
Note, the effects mentioned above can be described
by  estimating $\Delta \as= \as(f+1)-\as(f) \approx d\as(f)/df$ :
$$\Ds \Delta \as \approx \langle
\left(-{\partial_f \cM_n(a,f,Q_{exp})\over
\partial_a\cM_n(a,f,Q_{exp})}\right)
\rangle $$
Here, the brackets $ \langle (...)\rangle$ denote the average over
experimental values of $Q^2_{exp}$.
At the one--loop level, this expression leads to the simple estimate
$$
{\Delta \as \over \as} \sim
 -\Delta\beta_{(1)} \as \langle (\ln(\frac{Q_{exp}}{Q_0}))\rangle,
$$
in qualitative agreement with the results in Table 1.
\item The final results for $\alpha_{s}(M_Z)$  depend on the $Q^2_0$
choice.
 For the $f$-fixed scheme this dependence is {\bf not smaller} than
$2 \%$ (two thousandth of the absolute value)
and for the spline scheme, it is {\bf within only} $0.5 \%$
of the value of $\as(M_Z)$.
Of course, the value of $\alpha_{s}^{spl}(Q_0)$ lies
between the corresponding values of $\alpha_{s}^{\MSbar}(Q_0;f=4)$ and
$\alpha_{s}^{\MSbar}(Q_0;f=5)$, that reflects the real distribution
of the CCFR'97 experimental points. The suppression of the residual
dependence
on $Q_0^2$ for the spline scheme results gives an additional
``phenomenological" hint for preferring this scheme over the  $\MSbar$
$f$-fixed version.

\item The results of the fit are rather stable to the mass variations.
The $10 \%$ change of $M_c$ and $M_b$ yields less than $0.5\%$ change
for $\alpha_0$.
\end{itemize}

\vspace{2mm}
Comparing the spline and \MSbar f-fixed results we arrive at two main
conclusions:
\begin{enumerate}
\item
The spline scheme is more preferable than the traditional massless scheme,
to process the experimental data involving thresholds, the values of
$\as^{spl}(M_Z, Q_0)$ are focused tightly. The average value
$\alpha_{s}^{spl}{(M_Z)}=0.122$
is a little bit large than that the CCFR Collab. result for \asZ \,
obtained
recently in \cite{prep1} from the $Q^2$--evolution of $xF_3$ and $F_2$
for $Q^2> 5\,GeV^2$ and the invariant mass-squared of $W^2 > 10$ GeV$^2$
$$\asZ=0.119\pm 0.002\mbox{(exp.)}\pm 0.004\mbox{(theor)}.$$

\item
The threshold effects reveal themselves as approximately a $+1 \%$
correction \footnote{We have found the same estimate, $+1 \%$, for
the threshold
correction to \asZ for the processing the ``old" CCFR date
in \cite{CCFR93}} to the value of \asZ. The final results for the
average values of \asZ in the fit presented in Table 1 look like:
\ba
\alpha_{s}^{spl}{(M_Z)}&=&0.122\pm0.001~(theor)~\pm0.002~(stat) \nonumber
\\
\alpha_{s}^{\MSbar}{(M_Z;f=4)}&=&0.121\pm0.002~~(theor)~\pm0.003~(stat).
\nonumber
\ea
\end{enumerate}
Theoretical errors presented here include the uncertainties due to
Jacobi polynomial technique reconstruction and $Q_0^2$-deviation of the
$\as(M_Z)$ value in the fit.

\section{CCFR data and the light gluino window}
In Subsec. 4.2, we have obtained a comparatively small effect, one-two
thousandth to $\as(M_Z)$ for the threshold contribution at the fit.
The order of the effect is determined by the reason that only one
 $b$-quark threshold ``works" really in the region in question.
If Nature would provide few thresholds
in the experimental interval $3 \ GeV^2 \le Q^2 \le 200 \ GeV^2$,
then they combined influence in a fit becomes significant and the
preference of the spline scheme should look evident.
To demonstrate this here we took an attempt to reconcile the existence of
light  MSSM gluino ($\gl$) and the CCFR'97 data. The possibility
of the light gluino existence ($m_{\gl}$ \ is of the order $m_b$) was
intensively discussed few years ago in the context of the discrepancy
between low energy \as values and the LEP data at the $M_Z$ peak. This
discrepancy has a chance to be resolved by including the light gluino
~\cite{ENR-93}.

The Majorana gluino leads to large effects in the
evolution -- $\Delta\beta^{\gl}_{(1)} =2~$, \
$\Delta\beta^{\gl}_{(2)}=48~$
in Eq.(\ref{nMS}) and slows the running both the coupling constant, and
the moments $\cM_n(Q)$. This reinforcement of the contribution of a new
$\gl$ threshold must strongly influence the fit parameters.
Here we shall suggest the nucleon does not contain light gluino as a
parton,
and gluino reveals itself only in evolution law.
It is clear that the standard \MSbar -scheme at a fixed $f$ everywhere is
not adequate to the situation. We have performed the fit of CCFR'97 data
for
different values of the gluino mass $m_{\gl}$, the results of the fit for
\as at $m_{\gl}=2,~3,~4$ GeV and $m_{\gl}= 10$~GeV are presented in
Table 2.

Rather a strong growth of $\as^{spl(\gl)}(M_Z, Q_0)$ value
with the decrease of $m_{\gl}$, as well as a slight growth of $\chi^2$,
do not provide confidence that the data are
consistent with the gluino with mass $m_{\gl} \le 10$ GeV (see Table 2).
Let us consider the gluino mass
$m_{\gl}$ as a fit parameter and
``release" it. The best $\chi^2$ for \as is reached at $m_{\gl}^2$
\ba \la{mass} m_{\gl}^2 =103^{-50}_{+\infty} \ GeV^2,
~~\alpha_{s}^{spl(\gl)}{(M_Z)}=0.138,
\ea
with the asymmetrical stat. errors to the value of $m_{\gl}^2$,
$ (-50) / (+ \infty)$ GeV$^2$.
The latter means that the value of $m_{\gl}^2$ obtained in (\ref{mass})
is not reliable and
the CCFR'97 data ``push out" the light gluino to the
region of a more heavy gluino beyond the fitting region, \ie,
~$m_{\gl}^2 \ge 200 ~GeV$.
Moreover, the ``optimum" value of $ \alpha_{s}^{spl(\gl)}{(M_Z)}$
is over $3 \sigma$ higher than the world average value for \asZ
~\cite{Beth95}.

These results are in agreement with the conclusion in ~\cite{Fodor97} that
the light gluino is ruled out in others inclusive processes.

\section{Conclusion}
We have devised here a new approach for describing the two-loop
evolution of the moment $\cM_n(Q^2)$ of structure function of
lepton-nucleon DIS involving the threshold effects of heavy particles.
The approach employs an analytic quasi-exact two-loop RG solution
~\cite{tmf81,np92} within the framework of mass-dependent
remormalization group and on the results of paper~\cite{mikh}.

We adapt here the spline approximation (\ref{last}) to the above
mentioned RG solution and obtain as a result the simple modification
of the formulae of the standard massless \MSbar evolution.
For the particular case of a coupling constant evolution this
approximation effectively leads to the same result as the ``matching
condition".
Finally, this recipe provides a more realistic continuous description
for the $\cM_n(Q^2)$-evolution and looks rather simple from a practical
point of view.
We performed the processing of the modern CCFR data to extract
the value of $\as(Q)$ by  two different ways:

({\it i}) the traditional $\MSbar$-scheme at the fixed numbers of
flavors $f$;

({\it ii}) the spline scheme with break point at mass $M = m $;

The results for the \MSbar spline -- scheme processing of the data are the
most adequate to the situation both for the physical and practical
points of view. The threshold contribution to the value of $\alpha_s(M_z)$
consists of about $+1\%$; the extracted value of
 $\alpha_{s}^ {spl}{(M_Z)}$ is equal to $0.122\pm0.004$.
The order of the threshold effect is determined by the reason that only
one
b-quark threshold works really in the CCFR'97 experimental region.
We examine the possibility to reconcile the CCFR data and the MSSM light
gluino. The CCFR'97 data ``push out" the light gluino to the region of
more
heavy masses.
A reconciliation is possible for gluinos with mass $m_{\gl} \ge 14$ ~GeV,
but it seems that similar gluinos are less probable due to other
constraints (see ~\cite{ENR-93}).
\vspace{5mm}

\centerline{\bf Acknowledgements}
\vspace{2mm}
The authors are grateful to Dr. A. Kataev for fruitful discussions of
main results.
This investigation has been
supported in part by INTAS grant No 93-1180, two of us M.S.V. and S.A.V.
 were supported by the Russian Foundation for Fundamental
Research (RFFR) NN 96-02-17631 and 96-02-17435-a.

\begin{appendix}
\appendix
\section{Appendix}
Here we list the mass-dependent one-loop expressions for $I_1(z)$ and the
corresponding $\beta$ functions in the MOM  scheme. First, we write the
well-known exact expression for the fermion polarization operator
( $z= \mu^2/m_i^2$ )
\ba \la{exactH1}
\Ds I_{(1)}^q(z)=2 \sqrt{1+4/z}\left(1-{2\over z}\right)\ln\left(
\sqrt{1+z/4} + \sqrt{z/4}\right)+{4\over z} - {5\over 3}
\ \ ; \\
\la{asym1}
I_{1}^q(z\to \infty) = \ln z - (c_q = {5\over 3}) + O(1/z)\ \ ;\
I_{1}^q(z\to 0) = {z\over 5} + O(z^2)\ \ .
\ea
and the $\beta$ function
\ba
\beta_{(1)}= \frac{11}{3} C_A - \frac{4}{3} T_r\sum^{f}_{i=1}
\left(1-h_i(z) \right);  \nonumber \\
\mbox{where} ~~h_i(z)=h^{q}_i(z) = \frac{6}{z} -
\frac{12}{z^{3/2}} \frac{1}{\sqrt{1+z/4}}
      \ln\left(\sqrt{1+z/4} + \sqrt{z/4}\right)
\ea
The three--gluon contribution  $I^g_0(z)$ can be expressed in
terms of dilogarithms $Li_2(z)$, the final formula for $I^g_0(z)$
is too cumbersome and we do not demonstrate it here. As it was
predicted in ~\cite{NaWe}, the result for the corresponding
three--gluon contribution to the $\beta$-function, $h_i(z)=h^g_i(z)$
--function may be expressed in terms of  dilogarithms, as well,
$$
\baa
\la{3-g}
\Ds h^{g}_i(z) = \frac{18}{z} - \frac{36}{z^{3/2}} \frac{1}{\sqrt{1+z/4}}
      \ln\left(\sqrt{1+z/4} + \sqrt{{z/4}}\right) - &&  \\
\Ds - 6\Biggl\{\frac{1}{\sqrt{z}(1+z/3)} \frac{1}{\sqrt{1+z/4}}
  \ln\left(\sqrt{1+z/4}+\sqrt{{z/4}}\right) + && \nonumber \\
\Ds + \frac{1}{\sqrt{3}z} \mbox{\bf Im}
\left[Li_2(z_3^{\ast})-Li_2(z_3 )+Li_2(z_4^{\ast})-Li_2(z_4)\right]
\Biggr\}, && \nonumber  \\
\Ds z_3=-\left( 1+\frac{i}{\sqrt{3}}\right)
 \left( \frac{\sqrt{z}}{2\sqrt{1+z/4}+i\sqrt{z/3}}\right);\ \
z_4=\left(1-\frac{i}{\sqrt{3}}\right)
\left( \frac{\sqrt{z}}{2\sqrt{1+z/4}-i\sqrt{z/3}}\right).&&
\eaa
$$
The function $h^{g}_i(z)$  may be described by the rational approximation
~\cite{NaWe}:
\ba
h^{g}_i(z) \to \tilde{h}^{g}_i(z) = \frac{(1.2z+1)z}{(0.15z+1)(0.4z+1)}.
\ea
This approximation works well (better than $1\%$ of accuracy) when $z >
1$,
but when $z \approx 10^{-2} - 10^{-3}$, the accuracy is about $10\%$.
One can restore the corresponding approximate expression for
$I^g_0(z)$ by the elementary integration of $\tilde{h}^{g}(z)$
\ba   \la{a5}
\tilde{I_1^g}(z) &=&A\cdot \ln(1+z/z_1)+(1-A)\cdot \ln(1+z/z_2)~; \\
A&=&21/10 ~ ;\ \ z_1 = 20/3~;\ \ z_2 = 5/2~; \nonumber  \\
\la{a6}
\tilde{I_1^g}(z \to \infty) &=& \ln z - (c_g \approx 2.98) + O(1/z) \ \
;\ea
\end{appendix}

\newpage
{ {
\begin{tabular}{|l||l|c|c|} \hline
Quark      &           &2-loops         &                \\  \cline{2-4}
content    & $\chi^2 $ &$\alpha_0$      &  $\as(M_Z)$ \ \   \\  \hline
\multicolumn{4}{|c|}{ $Q_0^2=3 ~GeV^2$ }                 \\  \hline
\multicolumn{4}{|c|}{  \MSbar -- scheme, $f$ -- fixed, $M_q=0$ }\\  \hline
   uds   & 101.2       &0.3674$\pm$ 0.014& 0.1233$\pm$ 0.0015 \\ \hline
   udsc  & 102.9      &0.3545 $\pm$0.0220& 0.1219$\pm$0.0020 \\ \hline
   udscb & 105.1      &0.3411$\pm$ 0.0495& 0.1204 $\pm$0.0050  \\ \hline
\multicolumn{4}{|c|}{ Spline,\ $ M_c=1.3$  GeV, $ M_b= 5$ GeV} \\  \hline
uds+(c+b)& 103.5       &0.3494$\pm$0.0180& 0.1214 $\pm$0.0020    \\ \hline
\multicolumn{4}{|c|}{ $Q_0^2=30 ~GeV^2$ }                 \\  \hline
\multicolumn{4}{|c|}{  \MSbar -- scheme, $f$ -- fixed, $M_q=0$ }\\  \hline
   uds   & 104.5        &0.2156$\pm$0.0084 &0.1206 $\pm$0.0030 \\ \hline
   udsc  & 105.5        &0.2183$\pm$0.0080 &0.1214 $\pm$0.0030 \\ \hline
   udscb & 106.7        &0.2210$\pm$0.0080 &0.1222 $\pm$0.0030  \\ \hline
\multicolumn{4}{|c|}{ Spline,\ \ $ M_c=1.3$  GeV, $ M_b= 5$ GeV} \\ \hline
uds+(c+b)& 105.9       &0.2195$\pm$ 0.008& 0.1218 $\pm$0.0025    \\ \hline
\multicolumn{4}{|c|}{ $Q_0^2=200 ~GeV^2$ }                \\  \hline
\multicolumn{4}{|c|}{  \MSbar -- scheme, $f$ -- fixed, $M_q=0$ }\\  \hline
   uds   & 103.3       &0.1632$\pm$0.0046&0.1173 $\pm$0.0030    \\ \hline
   udsc  & 104.0       &0.1681$\pm$0.0050& 0.1197 $\pm$0.0025  \\ \hline
   udscb & 104.9       &0.1732$\pm$0.0048&0.1222 $\pm$0.0020  \\ \hline
\multicolumn{4}{|c|}{ Spline,\ \ $ M_c=1.3$  GeV, $ M_b= 5$ GeV}\\  \hline
uds+(c+b)& 104.7       &0.1721$\pm$00050 & 0.1217 $\pm$0.0020   \\ \hline
\end{tabular}
 }} \\
\vspace{1cm}

  {\bf{ Table 1 }}
The results of the NLO QCD fit in the $\MSbar$-scheme of the New
CCFR  \cite{prep1} $xF_3$ structure function data in a wide kinematical
region: $0.0075\leq x \leq 0.75$ and $3$ GeV$^2 < Q^2 < 200$ GeV$^2$
($N_{exp. p.}=102$). The value of the coupling constant is determined
for different numbers of the flavor (in the up-parts of Table for every
$Q_0^2$) and for few values of the momentum transfer
$\alpha_0=\alpha(Q_0^2=3,\ \ 30,\ \ 200 \ GeV^2)$. We present in the first
column the flavor content involved in QCD--evolution using in fit.
The results of fit with the matching at the thresholds corresponding to
$m_c=1.3$ ~GeV and $m_b=5$ ~GeV are presented in the down-parts (for every
$Q_0^2$) of the Table.

\newpage
{\large \bf New CCFR data and the light gluino window}

\vspace{5mm}
{ {
\begin{tabular}{|c||l|c|c|} \hline
Particle             &       &\multicolumn{2}{|c|}{2-loops} \\ \cline{2-4}
content              & $\chi^2       $ &  $\alpha^{spl(\gl)}_0$   &
$\as^{spl(\gl)}(M_Z)$ \ \  \\  \hline
\multicolumn{4}{|c|}{ $Q_0^2=3 GeV^2$,\ \ $ M_c=1.3$  GeV,
$ M_b= 5$ GeV, }                 \\  \hline
 Spline &\multicolumn{3}{|c|}{  $ M_{\tilde{g}}=2$ GeV}\\  \hline
 uds+(c+b+$\tilde{g}$) &111  &0.3102$\pm$0.015&0.1446 \\ \hline
 Spline &\multicolumn{3}{|c|}{  $ M_{\tilde{g}}=3$ GeV}\\  \hline
uds+(c+b+$\tilde{g}$) &109  &0.3193$\pm$0.012&0.1428\\ \hline
 Spline &\multicolumn{3}{|c|}{  $ M_{\tilde{g}}=4$ GeV}\\  \hline
 uds+(c+b+$\tilde{g}$)&107&0.3260$\pm$0.019&0.1415 \\ \hline
 Spline &\multicolumn{3}{|c|}{  $ M_{\tilde{g}}=10$ GeV}\\  \hline
uds+(c+b+$\tilde{g}$) &103&0.3475$\pm$0.018&0.1372\\ \hline
\end{tabular}
 }} \\

\vspace{5mm}

  {\bf{ Table 2 }}
The results of fit with the matching at the thresholds corresponding to
$m_c=1.3 ~GeV$, $m_b=5 ~GeV$ and $m_{\tilde{g}}=2,\ 3,\ 4,\ 10 ~GeV$ are
presented at the  Table 2.
The value of $\as^{spl(\gl)}(M_Z)$ is calculated with the error about
$\pm 0.004$.

\begin{thebibliography}{99}
\bibitem{yuri} Yu.L. Dokshitzer and D.V.Shirkov,                   
     {\it Z.Phys.C} {\bf 67} (1995) 449.
\bibitem{mikh} D.V. Shirkov and  S.V. Mikhailov,                     
        {\it Z.Phys.C} {\bf 63} (1994) 463.
\bibitem{chyla} J. Chyla
{\it Phys. Lett.} {\bf B351} (1995)  325;                            
A.L.Kataev et al.,
{\it Phys. Lett.} {\bf B388} (1996)  179;                            
\bibitem{kataev} A.L. Kataev et al., Preprint JINR E2-97-194, Dubna;
INR-947/97; hep-ph/9706534.
\bibitem{ST95} A.V. Sidorov, M.V. Tokarev
{\it Phys. Lett.} {\bf B358} (1995) 353  .                            
\bibitem{bern} W.Bernreuther and W.Wetzel, \np{197}{228}{82}.         
\bibitem{marc} W. Marciano, \pr{29}{580}{84}.                         
\bibitem{GA} A. Gonzales-Arroyo, C. Lopez, F.J. Yndurain
\np {153} {161} {79}.            
\bibitem{gotts} B.J. Edwards and T.D. Gottschalk, \np{196}{328}{82}.  
\bibitem{tmf81} D.V. Shirkov, {\it Theor. Mat. Phys. (USA)}            
                {\bf 49} (1981) 1039.
\bibitem{np92} D.V. Shirkov, \np {371} {267} {92} .                    
\bibitem{Brussel95} D.V. Shirkov, JINR Preprint E2-95-341, Dubna.      
\bibitem{LRV}
S.A.~Larin, T.~van Ritbergen and J.A.M.~Vermaseren,
{\em  Nucl. Phys.} {\bf  B427} (1994) 41;
S.A.~Larin, P. ~Nogueira, T.~van Ritbergen and J.A.M.~Vermaseren,
NIKHEV-96-010, hep-ph/9605317.
\bibitem{tbook}  F.J. Yndurain, Sec. 23 in "Quantum Chromodynamics.
An Introduction to the Theory of Qarks and Gluons"( Springer, 1983).  
\bibitem{GePo} H. Georgi and H.D. Politzer, \pr{14}{1829}{76}.         
\bibitem{NaWe}  O.Nachtmann, and W. Wetzel, \np{146}{273}{78}.         
\bibitem{mr85}  S.V. Mikhailov and A.V. Radyushkin   \np{254}{89}{85}.
\bibitem{japan} T.Yoshino, K.Hagiwara, {\it Z.Phys.C} {\bf 24} (1984) 185.
\bibitem{CCFR93}
CCFR Collab., W.C.Leung et al.,
{\it Phys. Lett.} {\bf B317}, 655 (1993);
CCFR Collab., P.Z.Quintas et al.,
{\it Phys. Rev. Lett.} {\bf 71}, 1307 (1993);
\bibitem{prep1}
W.~G.~Seligman,
NEVIS-report-292, Jan 1997;
CCFR-NuTeV Collab., W.~G.~Seligman et al.,
 hep-ex/9701017.
\bibitem{Kriv}
V.G.Krivokhizhin et al.,
{\it Z. Phys.} {\bf C36},51 (1987); {\bf C48}, 347 (1990). 
\bibitem{Jacobi}
G.Parisi, N.Sourlas,
{\it Nucl. Phys.} {\bf B151} (1979) 421;
 I.S.Barker, C.B.Langensiepen and G.Shaw,
{\it Nucl. Phys.} {\bf B186} (1981) 61.
\bibitem{KaSi}
A.L.Kataev and A.V.Sidorov,
{\it Phys. Lett.} {\bf B331} (1994) 179.
\bibitem{GEORGI}
 O. Nachtmann, {\it Nucl. Phys.} {\bf B 63} (1973) 237;
H.~Georgi, H.D.~Politzer, {\it Phys. Rev.} {\bf D 14} (1976) 1829;
S. Wandzura, {\it Nucl. Phys.} {\bf B 122} (1977) 412.
\bibitem{SSM96}
D.V.~Shirkov,  A.V.Sidorov, S.V.~Mikhailov,  JINR Preprint E2-96-285,
Dubna; hep-ph/9607472.
\bibitem{Beth95}
S. Bethke,
{\it Nucl. Phys. (Proc. Suppl.)} { \bf 39 B, C}   (1995) 198.
\bibitem{ENR-93} J. Ellis, D.V. Nanopoulos, D.A. Ross, \pl{305}{375}{93};
 U.Amaldi, W. de Boer and H.Furstenau, \pl{260}{447}{91};
 I. Antoniadis, J. Ellis, D.V. Nanopoulos,
		{\it Phys.Lett. B}{\bf 262} (1991) 109;
 M. Jezabek and J. Kuhn, {\it Phys.Lett.}{\bf B 301}
		(1993) 121;
 J. Ellis, D.V. Nanopoulos, D.A. Ross, \pl{305}{375}{93}.
\bibitem{Fodor97}
F. Csikor and Z. Fodor, {\it Phys.Rev.Lett.} {\bf 78} (1997) 1335.
\end{thebibliography}
\end{document}